\documentclass[titlepage,12pt]{article}

\usepackage{epsfig,pslatex}

\newcommand{\be}{\begin{equation}}
\newcommand{\ee}{\end{equation}}

\newcommand{\bea}{\begin{eqnarray}}
\newcommand{\eea}{\end{eqnarray}}
\newcommand{\ttbar}{t {\bar t}}
\newcommand{\mtt}{M_{t \bar t}}

\begin{document}
\begin{titlepage}
\vspace{0.01cm}
\begin{center}
{\LARGE {\bf Electroweak corrections to 
$t\bar{t}$ production at hadron colliders
}} \\
\vspace{2cm}
{\large{\bf W. Bernreuther\,$^{a,}$\footnote{Email:
{\tt breuther@physik.rwth-aachen.de}},
M. F\"ucker \,$^{a,}$\footnote{Email:
{\tt fuecker@physik.rwth-aachen.de}},
Z. G. Si\,$^{b,}$\footnote{Email: {\tt zgsi@sdu.edu.cn}}\footnote{
Speaker at
{\it International Workshop on Top Quark Physics},
La Biodola, Elba, Italy, 18-24 May 2008.
}
}}
\par\vspace{1cm}
$^a$Institut f\"ur Theoretische Physik, RWTH Aachen, 52056 Aachen, Germany\\
$^b$Department of Physics, Shandong University, Jinan, Shandong
250100, China\\
\par\vspace{1cm}
{\bf Abstract}\\
\parbox[t]{\textwidth}
{We report on our recent work on electroweak corrections
to $t\bar{t}$ production at hadron colliders. 
Specifically, we discuss the weak-interaction 
contributions to the 
top quark transverse momentum and $t \bar{t}$ invariant 
mass distributions and an induced parity-violating 
 top-spin asymmetry.
}
\end{center}
\vspace*{2cm}
PACS number(s): 12.15.Lk, 12.38.Bx, 13.88.+e, 14.65.Ha
\end{titlepage}

\setcounter{footnote}{0}
\renewcommand{\thefootnote}{\arabic{footnote}}
\setcounter{page}{1}


LHC is planned to start operation within this year.
Once LHC will run near design luminosity, 
huge number of $t\bar{t}$ events will be produced.
 This will allow detailed  exploration of the  properties of top quarks. 
For this aim, theoretical predictions related to top quark production and
decay should be made as precisely as possible within the SM. 
 Predictions for unpolarized $t \bar t$
production have long been known at  next-to-leading order (NLO) QCD 
\cite{Nason:1987xz},
and these NLO results were
refined by resummation of soft gluon and threshold logarithms
\cite{Bonciani:1998vc}.
Moreover, $t \bar t$ production and decay including the full spin
degrees of freedom of the intermediate $t$ and $\bar t$ resonances 
were determined to NLO QCD some time ago 
\cite{Bernreuther:2000yn,Bernreuther:2004jv}.
Spin effects of top quarks can be reliably predicted in view of  the short-distance nature of their
interactions, and are expected to play an important role in refined data analysis.

A complete NLO analysis of $t \bar t$ production within  the SM should include
also the electroweak radiative corrections.  Though 
they turn out to be marginal  for the production cross section $\sigma_{t
  \bar t}$ at the Tevatron and at the LHC,
they may be important for distributions at large transverse top-quark
momentum or large $t \bar t$ invariant mass, due to large Sudakov
logarithms \cite{Melles:2001ye}.
Moreover, the weak interactions induce
small parity-violating effects, and for full exploration and interpretation
of future data it is important to obtain definite SM
predictions also for these effects. 
Weak interaction corrections to  hadronic $t \bar t$ production have been
studied  in a number of papers. The order
$\alpha_s^2\alpha$ weak-QCD
corrections to $q {\bar q} \to t \bar t$ and $gg \to t \bar t$ 
 were analyzed  in \cite{Beenakker:1993yr} (c.f. also \cite{Kao:1997bs}).
For $q \bar q \to t \bar t (g)$, full determinations  of these  corrections, including the
infrared-divergent box contributions and the corresponding real gluon
radiation, were made in 
\cite{Bernreuther:2005is,Kuhn:2005it}. The 
order $\alpha_s^2\alpha$ corrections to $gg \to t \bar t$ including
the quark triangle diagrams $gg \to Z \to t \bar t$ were investigated
in \cite{Bernreuther:2006vg,Kuhn:2006vh,Moretti:2006nf}, 
and additional contributions in  \cite{Bernreuther:2008md}. The purely
photonic corrections were recently also calculated \cite{Hollik:2007sw}.
 Parity violation in $t \bar t$ production was
analyzed within the SM
in \cite{Kao:1997bs,Bernreuther:2005is,Bernreuther:2006vg,Bernreuther:2008md,Kao:1994rn,Kao:1999kj,
Beccaria:2005jd}. Investigations of non-SM effects include
 \cite{Kao:1999kj,Li:1997gh}.


At hadron colliders top quark pairs are produced predominantly by 
the strong interactions. The QCD corrections for the  
subprocesses $i \to t {\bar t}+X$, $(i=q {\bar
q}, gg, gq, g {\bar q})$  are known to order $\alpha_s^3$. As
mentioned above,  the leading corrections involving electroweak interactions 
are also available.
We shall focus  here on weak-interaction contributions, 
but shall mention for
completeness also QED corrections.
For the partonic process 
\begin{equation}
g + g \to t + \bar{t},
\label{gg1}
\end{equation}
the leading electroweak contributions are of order $\alpha \alpha_s^2$,
while for 
\begin{equation}
q + \bar{q} \rightarrow t + \bar{t}, \qquad q \neq b,
\label{qq1}
\end{equation}
there  are  
the order  $\alpha^2$  Born contributions  (from 
$q \bar q \to \gamma, Z \to t \bar t$) and 
the mixed QCD electroweak corrections of order  $\alpha_s^2 \alpha$.
  Due to color conservation there are no  order
$\alpha_s \alpha$ interference terms for the s-channel amplitudes.
But for 
\begin{equation}
b\bar{b}\to t\bar{t}
\label{bb}
\end{equation}
 there is, at leading order,  a t-channel $W$ exchange contribution 
together with the s-channel
gluon, $Z$ and $\gamma$ exchange amplitudes. 
The t-channel term interferes with the QCD Born amplitude. Thus the leading electroweak 
contributions to (\ref{bb})
are of order $\alpha^2$ and $\alpha\alpha_s$. In view of the large
 parton luminosity for $qg$ scattering at the LHC,
  one should take into account also the reactions
\begin{equation}
gq  \to t\bar{t} q  \, , \qquad g \bar{q} \to t\bar{t} \bar{q} 
\label{gqqb}
\end{equation}
 and determine the weak corrections to the respective amplitudes.
Thus the NLO weak corrections to hadronic $t\bar{t}$ production
can be divided into three parts:  i)  the contributions of order
$\alpha^2$ and $\alpha\alpha_s$ to  the process (\ref{bb}), 
 ii) the contributions of order
$\alpha^2\alpha_s$ and $\alpha\alpha_s^2$ 
 to the processes (\ref{gqqb}), and
iii) the contributions of order
$\alpha^2$ and $\alpha\alpha_s^2$ to the reactions
(\ref{gg1}) and (\ref{qq1}).
In our analysis, we employ the 5-flavor 
scheme \cite{Aivazis:1993pi}, where the
(anti)proton is considered to contain also $b$ and $\bar b$ quarks in
its partonic sea. Thus the reaction  (\ref{bb}) is a leading-order
(LO) process in this
scheme, while  (\ref{gqqb}), for $q=b$,  
is a next-to-leading order correction
to  (\ref{bb}). As to the pure QED corrections which were
calculated in \cite{Hollik:2007sw}: it was pointed out in that work
that the dominant photonic corrections are due to  photon-gluon
 fusion, $\gamma + g \to t\bar{t}$.


\begin{table}[ht]
\caption{The $t\bar{t}$ cross section at the 
Tevatron ($\sqrt{s}=1.96TeV$) and at the LHC ($\sqrt{s}=14TeV$) in
units of pb, for
$m_t=172.7$ GeV, $m_H=120$ GeV and three different values of $\mu$ . We
put $\mu=\mu_R=\mu_F$. The numbers for the QCD and weak contributions
 were obtained using the NLO 
parton distribution functions CTEQ6.1M, while the NLO QED contributions 
 are from \cite{Hollik:2007sw} which used the PDF set \cite{Martin:2004dh}. }
\begin{center}
\begin{tabular}{|c|c|c|c|c|}\hline
\multicolumn{2}{|c|}{} & $\mu={m_t}/{2}$  & $\mu=m_t$  &
$\mu=2m_t$  \\ \hline
 Tevatron     & NLO QCD  &  7.493  & 7.105  & 6.314   \\ \cline{2-5}
  (pb)        & Weak / QED   & $0.0339$ & $0.0355$  &  $0.0346$ / $-0.102$ \\ \cline{1-5}
LHC          & NLO QCD & 868.150  & 850.385 & 793.543 \\ \cline{2-5}
(pb)         & Weak / QED  & $-14.127$  & $-10.790$ & $-8.368$ / $4.78$ \\ \cline{1-5}       
\end{tabular}
\end{center}
\end{table}


For $t\bar{t}$ production at the level of hadronic
collisions, the inclusive 
spin-summed $t\bar t$ cross section 
may be written, to NLO in the SM couplings, in the form
${\sigma} = {\sigma}^{(0)} + \delta{\sigma}^{(1)} +
\delta {\sigma}^{W} + \delta {\sigma}^{QED}$,  
where the first and second term are the LO (order $\alpha_s^2$) and
NLO (order $\alpha_s^3$) QCD contributions,  while the third and fourth
 term result from the electroweak corrections to  the processes 
 discussed above. Table~1 contains the contributions
from $gg \to t {\bar t} (g)$ and $q\bar{q}\to t\bar{t}(g)$ at NLO
QCD\footnote{In this table, we have omitted the order $\alpha_s^3$
  contributions from $qg$ and ${\bar q} g$ initial states, as this table
  serves only the purpose of exhibiting the relative size of the
  electroweak corrections.}, the weak
corrections of order $\alpha_s^2\alpha$ and $\alpha^2$, and the QED
corrections of order $\alpha_s^2\alpha$ and $\alpha^2\alpha_s$.
The numbers for ``weak'' in this table do not contain  the contribution from
the t-channel $W$-exchange in (\ref{bb}).
For the evaluation of
 the QCD and weak contributions  we have used $\overline{\rm MS}$
factorization and the  NLO parton distribution functions (PDF)
CTEQ6.1M \cite{Pumplin:2002vw}. The QED corrections are from
 \cite{Hollik:2007sw}, where DIS factorization and  the PDF
 set from \cite{Martin:2004dh} was used, which contain PDFs at NLO QCD and NLO QED. 
The table shows that the
weak correction to the total cross section 
is negative at the LHC and amounts to about $-1.3 \%$,
while it is about $0.5\%$ at the Tevatron. The  pure NLO QED
correction 
is about 0.6\% (-2\%) at the  LHC (Tevatron). 
Thus, the electroweak corrections to the cross section are much
smaller 
than the scale uncertainties 
of the fixed-order NLO QCD corrections. 

However, as we shall show below, for a number of distributions,
 which are among the key observables in the tool-kit  
for the  search of new physics in $t \bar t$ events, these  corrections 
do matter if one aims  at predictions with a precision
at the percent level.
\begin{figure}[ht]
\begin{center}
\includegraphics[width=6cm, height=6cm]{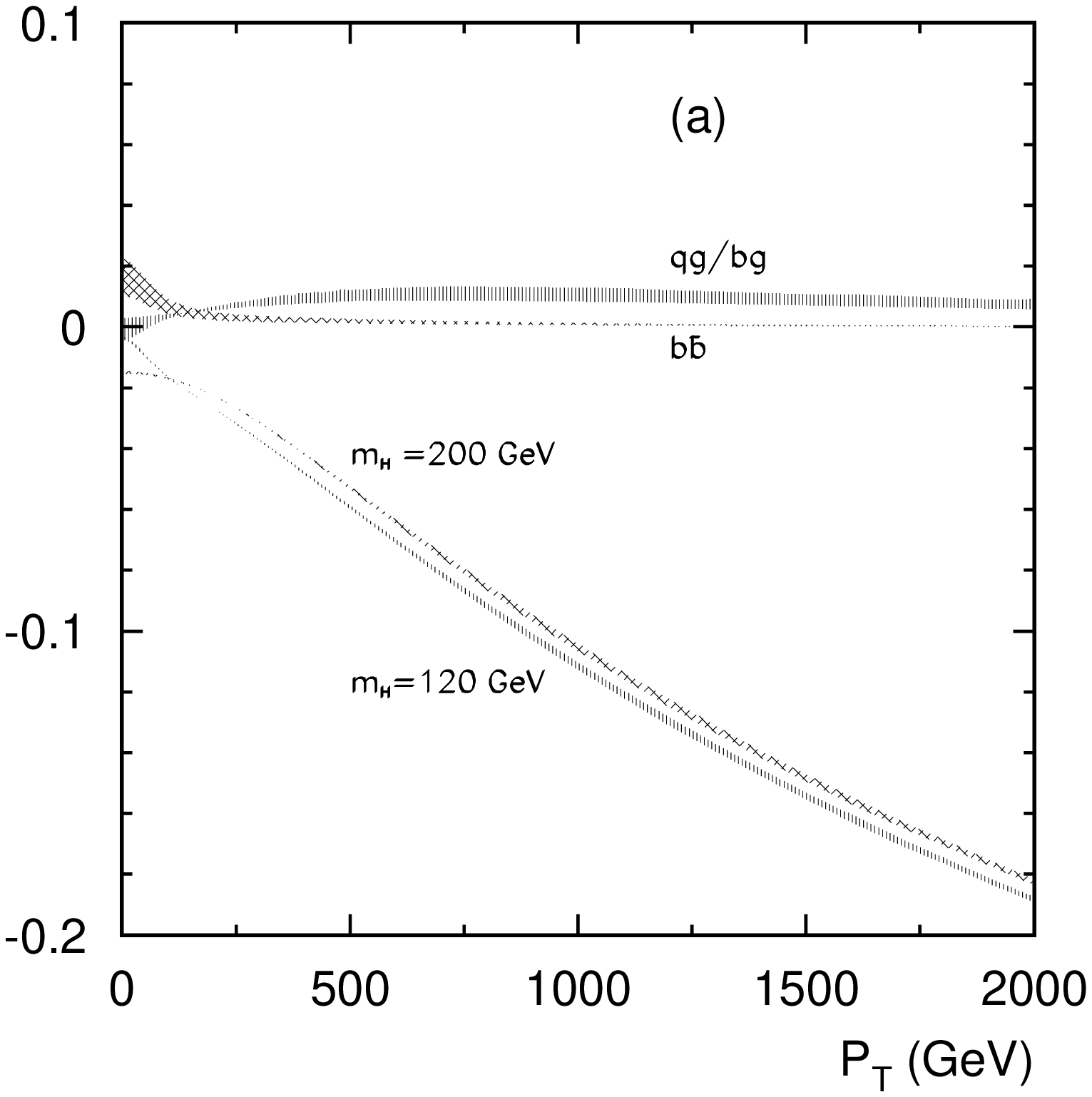}
\includegraphics[width=6cm, height=6cm]{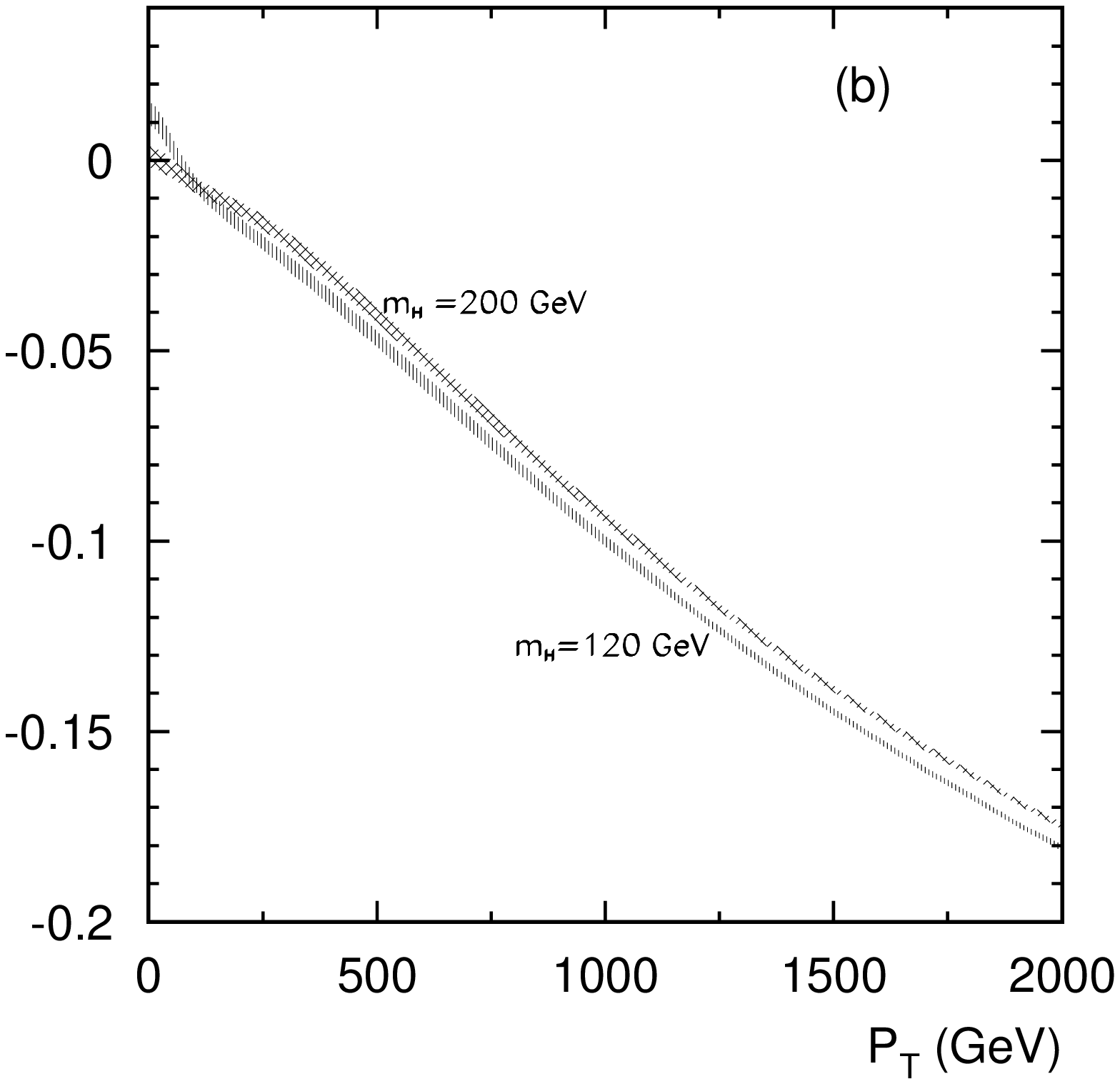}
\end{center}
\caption{ a) Ratios
$(d\sigma_{weak}/dp_T)/(d\sigma_{LO}/ dp_T)$, where $d\sigma_{weak}$  are
the weak-interaction corrections i), ii), and iii)
to the reactions (\ref{bb}), (\ref{gqqb}),
 and $q {\bar q}, gg \to \ttbar$ $(q\neq b)$, respectively. The latter
 corrections are shown for two different values of the Higgs boson
 mass. The hatched areas arise from scale variations as described in
 the text.  b) Sum of the ratios shown in a) for
two different values of $m_H$.}
 \label{fig:dpt}
 \end{figure}

For the computation of these distributions we use
 $m_t=172.7$ GeV,
$\alpha_s(2m_t)=0.1$, and  $\alpha(2m_t)=1/126.3$.
The LO QCD terms and the contributions of the corrections i) and iii) to the distributions
are evaluated with the  LO parton distribution functions (PDF)
CTEQ6.L1, while for the computation
 of the contributions from ii), which depend on the factorization
scale, the set CTEQ6.1M  \cite{Pumplin:2002vw} is used. 
The scale $\mu_F$ is varied between 
$m_t/2\leq\mu_F\leq 2 m_t$.  Dependence on the renormalization scale $\mu_R$
enters only via the ${\overline{\rm MS}}$ coupling $\alpha_s$. The ratio
of the corrections iii) and $d\sigma_{LO}$ is practically independent
of $\alpha_s$, while the corresponding ratios involving i) and ii)
vary weakly with $\mu_R$.

Fig.~\ref{fig:dpt}a shows the various 
weak-interaction contributions to the transverse
momentum distribution of the top quark at the LHC, normalized
to $d\sigma_{LO}/dp_T$. The hatched areas depict the range of values
when $\mu\equiv\mu_F=\mu_R$ is varied between $m_t/2$ and $2m_t$.
Fig.~\ref{fig:dpt}a  shows that the weak correction
 i) to the $p_T$ distribution of the top quark is positive and small. 
Its significance is confined to the region $p_T \leq$ 100 GeV,
where it dominates the
other weak corrections. However, in this region
these corrections make up only 
 between 1$\%$ and 2$\%$ of the  LO QCD $p_T$ distribution.
In the high $p_T$ region, where the weak-interaction corrections to
the $p_T$ spectrum become larger, the  contributions
from the processes (\ref{bb}) and (\ref{gqqb})
become less significant in comparison to the  weak
corrections iii).
Fig.~\ref{fig:dpt}b  displays the ratio of the sum of the weak corrections
i), ii), and iii) and the LO QCD contribution. The corrections are
 negative in almost the whole $p_T$ range. For large $p_T$ they are
 quite sizeable; for instance, for $p_T = 1000$ GeV they amount to
 $-10\%$ of the LO QCD contribution. The photonic corrections to
  the $p_T$ spectrum are also negative, but smaller in magnitude
  \cite{Hollik:2007sw}. For instance, at $p_T = 1000$ GeV they amount to
 $-2\%$ of $d\sigma_{LO}/dp_T$.

\begin{figure}[ht]
\begin{center}
\includegraphics[width=6cm, height=6cm]{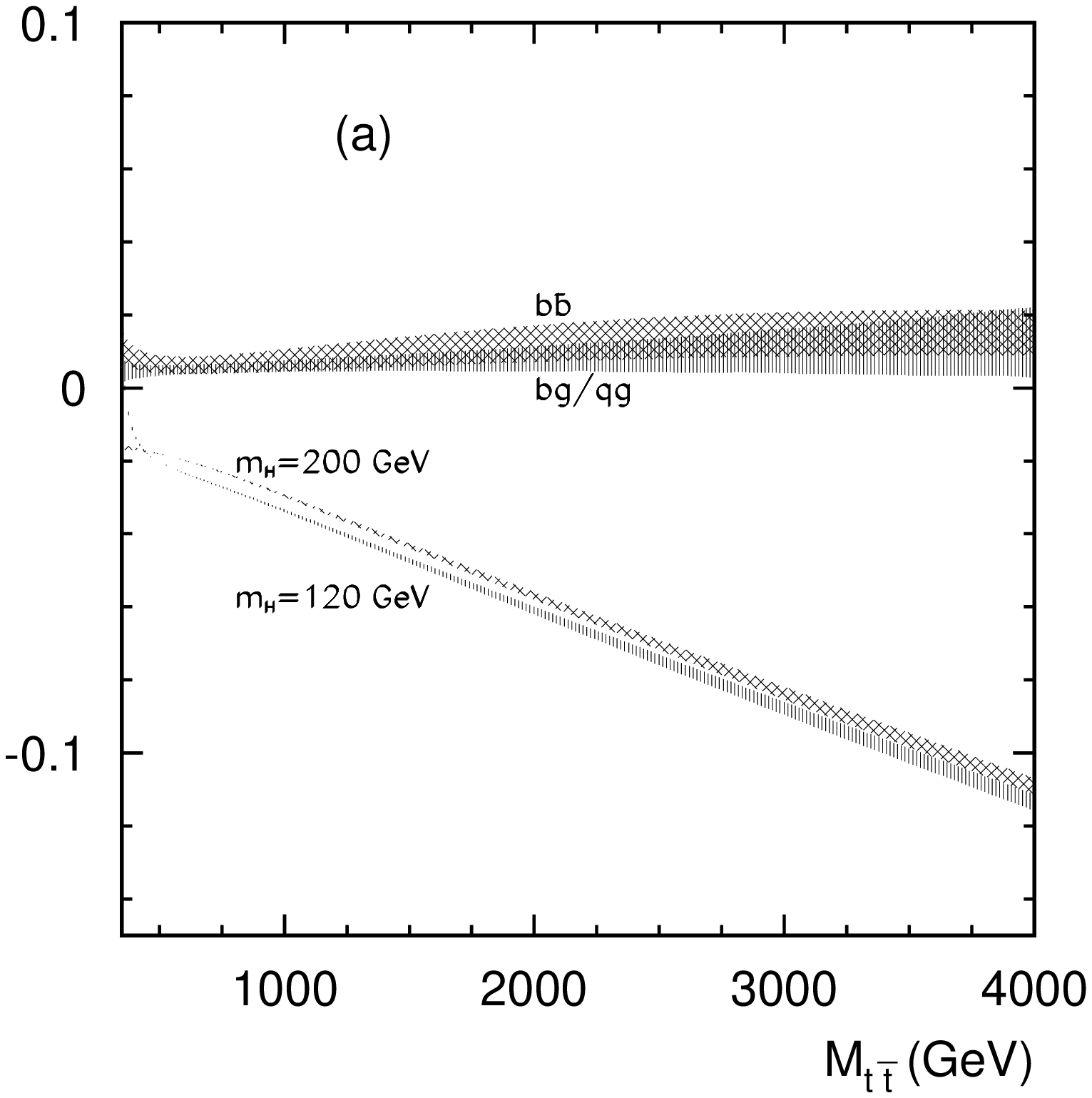}
\includegraphics[width=6cm, height=6cm]{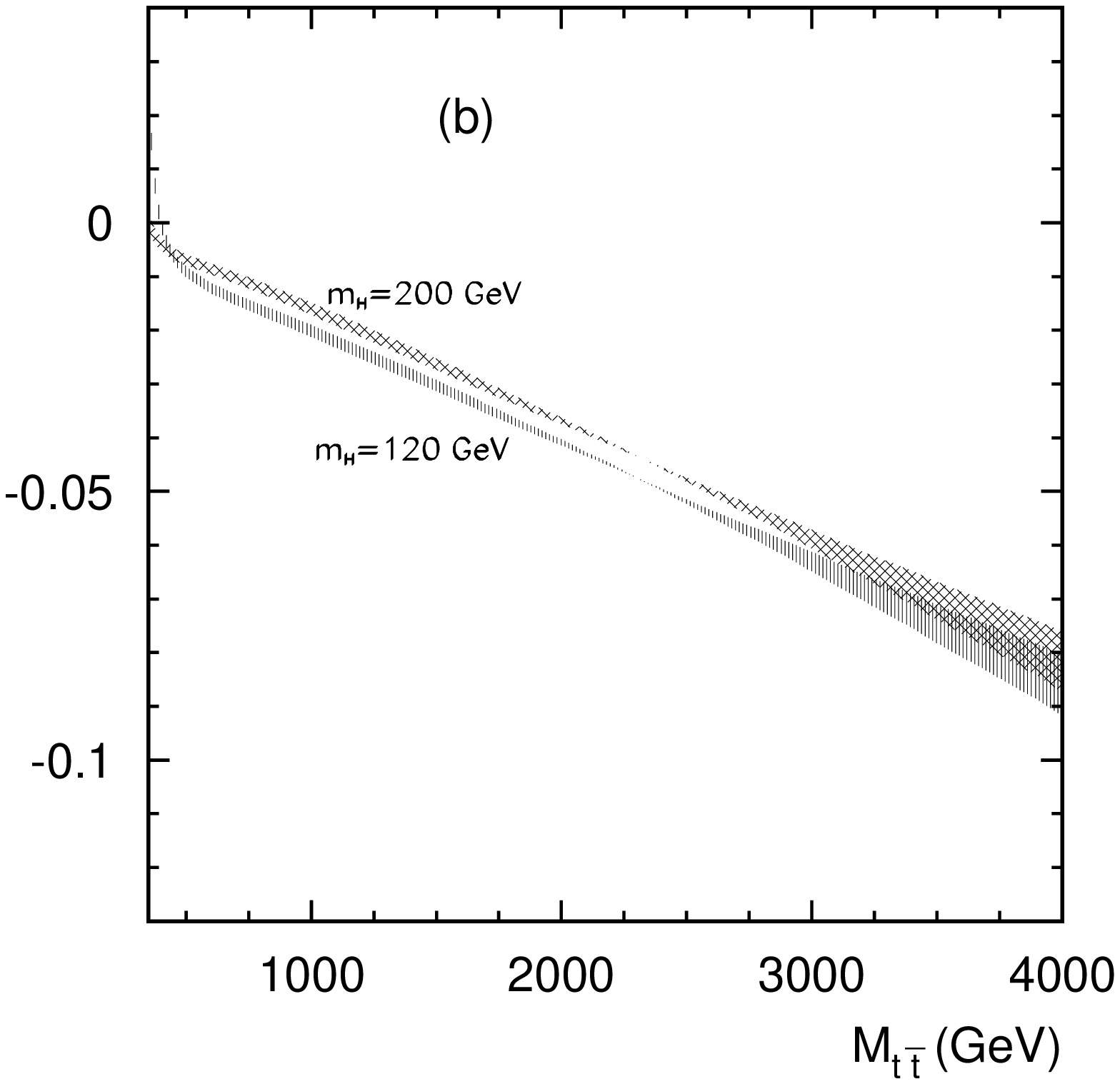}
\end{center}
\caption{ a) Ratios  $(d\sigma_{weak}/d\mtt)/(d\sigma_{LO}/d\mtt)$,
where $d\sigma_{weak}$  refers to
the weak-interaction corrections i), ii), and iii).
 The latter
 corrections are shown for two different values of the Higgs boson
 mass. The hatched areas arise from scale variations as described in
 the text.  b) Sum of the ratios shown in a)
for two different values of $m_H$.
}\label{fig:dmtt}
 \end{figure}

In Fig.~\ref{fig:dmtt}a the analogous ratios are displayed for the 
$M_{t \bar t}$ distribution. The weak-interaction corrections
i) and ii) are both positive and show a considerable scale uncertainty.
They reduce the magnitude of the leading weak corrections iii), which
are negative, as shown in
Fig.~\ref{fig:dmtt}b. 
Notice that in these ratios  the changes of $d\sigma_{weak}$
and $d\sigma_{LO}$ due to 
variations of the LO PDF and the LO QCD coupling with $\mu$ cancel to
a large extent. How large are the photonic corrections?
 The authors of \cite{Hollik:2007sw} have not computed the $M_{t \bar t}$
distribution, but the distribution of the partonic c.m. energy, and
obtained that the QED correction to this quantity is quite small
($\leq 1\%$ in magnitude) in most of the kinematic range.


Weak interaction-corrections  can induce also 
parity-violating (PV) effects.
One possibility to search for PV in hadronic $t\bar{t}$ pair
production  is to check whether the produced ensemble of $t$ and $\bar
t$ quarks  is longitudinally polarized. 
Let us discuss this possibility in the more general context
 of top-spin physics in $t\bar{t}$ pair production and decay.
The 
polarizations and spin-spin correlations which is
 imprinted upon the $t \bar t$ sample by the specific
 production dynamics lead, through
the parity-violating weak decays of these quarks, to characteristic
angular distributions and correlations among the final  state particles/jets.
 In
semileptonic top-quark decays the outgoing charged lepton 
is, according to the SM, the best top-spin
analyzer, while for non-leptonic top decays the resulting
least-energetic non-$b$ jet is a good and experimentally acceptable
choice \cite{Brandenburg:2002xr}.
Of the main
$t \bar t$ decay modes, that is, the all-jets, lepton + jets, and dilepton
channels, very probably only the latter two are useful for top-spin
physics, because the all-jets channels have rather low top-spin
 analyzer quality and large backgrounds.  Thus, for measuring top-spin
 effects in $t \bar t$ production and decay at the
Tevatron or LHC one may consider the reactions
\begin{equation}
p {\bar p}, \; p p \to t {\bar t} \, + X \to a({\bf p}_+) + {\bar
  b}({\bf p}_-) \, + X\, ,
\label{abreact}
\end{equation}
where $a$ and ${\bar b}$ denotes either a charged lepton ($\ell=e,\mu)$ or
a jet from $t$ and $\bar t$ decay, respectively, and ${\bf p}_+$ and 
${\bf p}_-$ denote the 3-momenta of these particles/jets in the
respective $t$ and $\bar t$ rest frame\footnote{For the lepton + jets and
for the dileptonic channels the
$t$ and $\bar t$ momenta, i.e., their rest frames can be kinematically
reconstructed up to ambiguities, which may be resolved with Monte Carlo
methods using the matrix element of the reaction.}. One may now 
choose two   polar vectors ${\bf{\hat a}}$ and  ${\bf{\hat b}}$
as reference axes, determine the angles 
$\theta_+ =\angle({\bf p}_+,{\bf{\hat a}})$ and
$\theta_- =\angle({\bf p}_-,{\bf{\hat b}})$ event by event, and 
consider the double distribution
\begin{equation}
\frac{1}{\sigma_{ab}}\frac{d\sigma}{d\cos\theta_+ d\cos\theta_-}
= \frac{1}{4}\left(1+ B_+\cos\theta_+ +  B_-\cos\theta_- -
C \cos\theta_+ \cos\theta_- \right) \, ,
\label{ddistab}
\end{equation}
where $\sigma_{ab}$ is the cross section of the channel (\ref{abreact}).
The right-hand side of (\ref{ddistab}) is the a priori form of this 
distribution if no cuts are applied. In the  presence of cuts the
shape of the distribution will in general 
be distorted. Nevertheless, one may use
the bilinear form (\ref{ddistab}) as an estimator in fits to data.
The coefficient $C$ contains the information about the parity-even
$t \bar t$ spin correlations. 
These distributions were predicted for the Tevatron and the LHC 
in \cite{Bernreuther:2004jv} to  NLO
QCD for an number of reference axes.
It is straightforward to add to these NLO
QCD results the weak interaction corrections which
may be enhanced by suitable cuts on $\mtt$. Detailed results can be
found in \cite{Bernreuther:2005is,Bernreuther:2006vg}.

The PV dynamics that contributes
to  $t\bar t$ production leads to a polarization of the $t$ and $\bar
t$ samples along some polar vector, i.e., to non-zero expectation
 values $<{\bf S}_t\cdot {\bf \hat{a}}>$, $<{\bf S}_{\bar t}\cdot {\bf
   \hat{b}}>$. The information about these  parity-odd 
 (anti)top-spin effects 
is contained in the coefficients $B_{\pm}$
of (\ref{ddistab}). The highest sensitivity to such effects
is achieved  when one uses the charged lepton from semileptonic
$t$ or $\bar t$ decay as top-spin analyzer. Thus, we consider the
 reactions 
\begin{equation}
p {\bar p}, \; p p \to t {\bar t} \, +  X \to \ell^+({\bf p}_+) \, +   X\, ,
\label{ellreact}
\end{equation}
where $\ell=e, \mu$. (Experimentally, the event
selection should use $b$-tagging, etc. in order to discriminate against
single $t$ production, which also contributes to
the  final state (\ref{ellreact}).)
 Integrating (\ref{ddistab}) with respect to
$\cos \theta_-$ yields the distribution 
\begin{equation}
\frac{1}{\sigma_{\ell}}\frac{d\sigma}{d\cos\theta_+}
= \frac{1}{2}\left(1+ B_+\cos\theta_+ \right) \, .
\label{ddispvl}
\end{equation}
As to the choice of the reference axes 
 ${\bf{\hat a}}$ and  ${\bf{\hat b}}$, 
we consider here the helicity basis
 \cite{Bernreuther:2004jv}, which is
the best choice for the LHC. (For the Tevatron, a suitable choice
 would be the beam basis.)
The distribution (\ref{ddispvl}) leads to  the
parity-violating asymmetry
\begin{equation}
A_{PV} \equiv \frac{N_+ - N_-}{N_+ + N_-} = \frac{B_+}{2} \, ,
\label{asapv}
\end{equation}
where $N_\pm$ is the number of events (\ref{ellreact}) with
$\cos\theta_+$ larger or smaller than zero. 

For the computation of $B_+$ we need several
 ingredients. For brevity, we consider here $A_{PV}$ 
 only for the LHC. Here $t \bar t$ production is dominated by gluon fusion,
 and the top quarks are on average (moderately) relativistic.
Thus, as already said above, choosing the top-quark direction
 of flight as the ``top-spin quantization axis'' approximately 
 optimizes  $A_{PV}$
 for a given PV production dynamics. At the level of the intermediate 
 $t \bar t$ events, the basic PV spin-asymmetry is the 
helicity asymmetry for $t$ quarks, $Z_{hel}$:
\begin{equation}
Z_{hel}\, = \,  \frac{d\sigma_{+}}{d\mtt} - 
\frac{d\sigma_{-} }{d\mtt}  \, , \quad {\rm and} \quad
\Delta_{hel}\, = \, \frac{Z_{hel}}{d\sigma_{LO}/d\mtt}\, .
\label{helasy}
\end{equation}
The subscripts $\pm$ in (\ref{helasy}) refer to a  $t$ quark with
positive/negative
helicity while the helicity states of the $\bar t$ are summed. 
The asymmetry coefficient $B_+$ is given by
\cite{Bernreuther:2006vg}
\begin{equation}
B_+ = \kappa_+ \frac{\int dM_{t\bar t} \, Z_{hel}(M_{t\bar
    t})}{\sigma_{t\bar t}} \, ,
\label{eqforbp}
\end{equation}
where  $\kappa_+$
is the top-spin analyzing power of $\ell^+$. In the SM
$\kappa_+=1$ to lowest order and $\kappa_+=0.9984$ including the order
$\alpha_s$ QCD corrections. 
\begin{figure}[ht]
\begin{center}
\includegraphics[width=6cm, height=6cm]{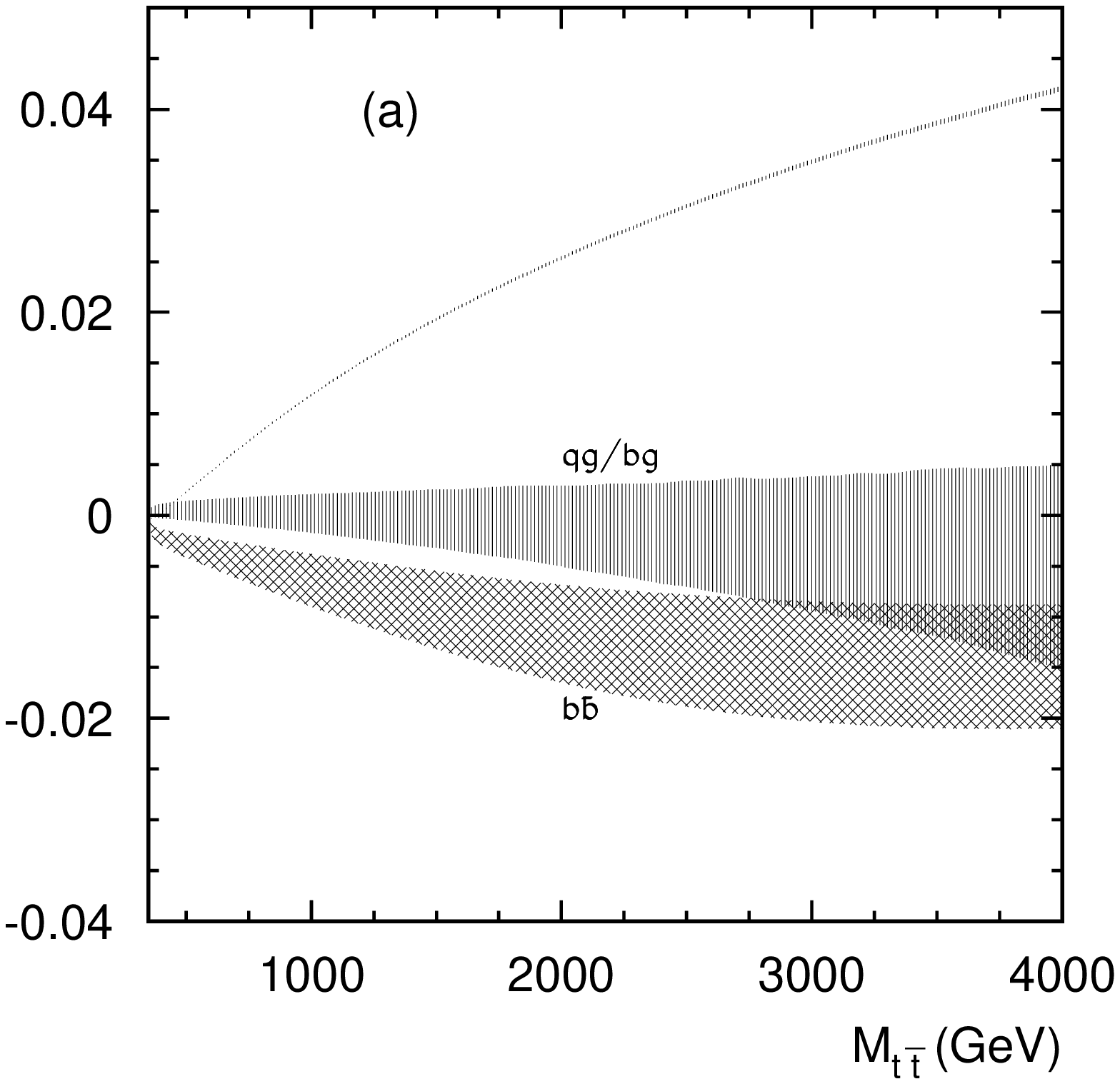}
\includegraphics[width=6cm, height=6cm]{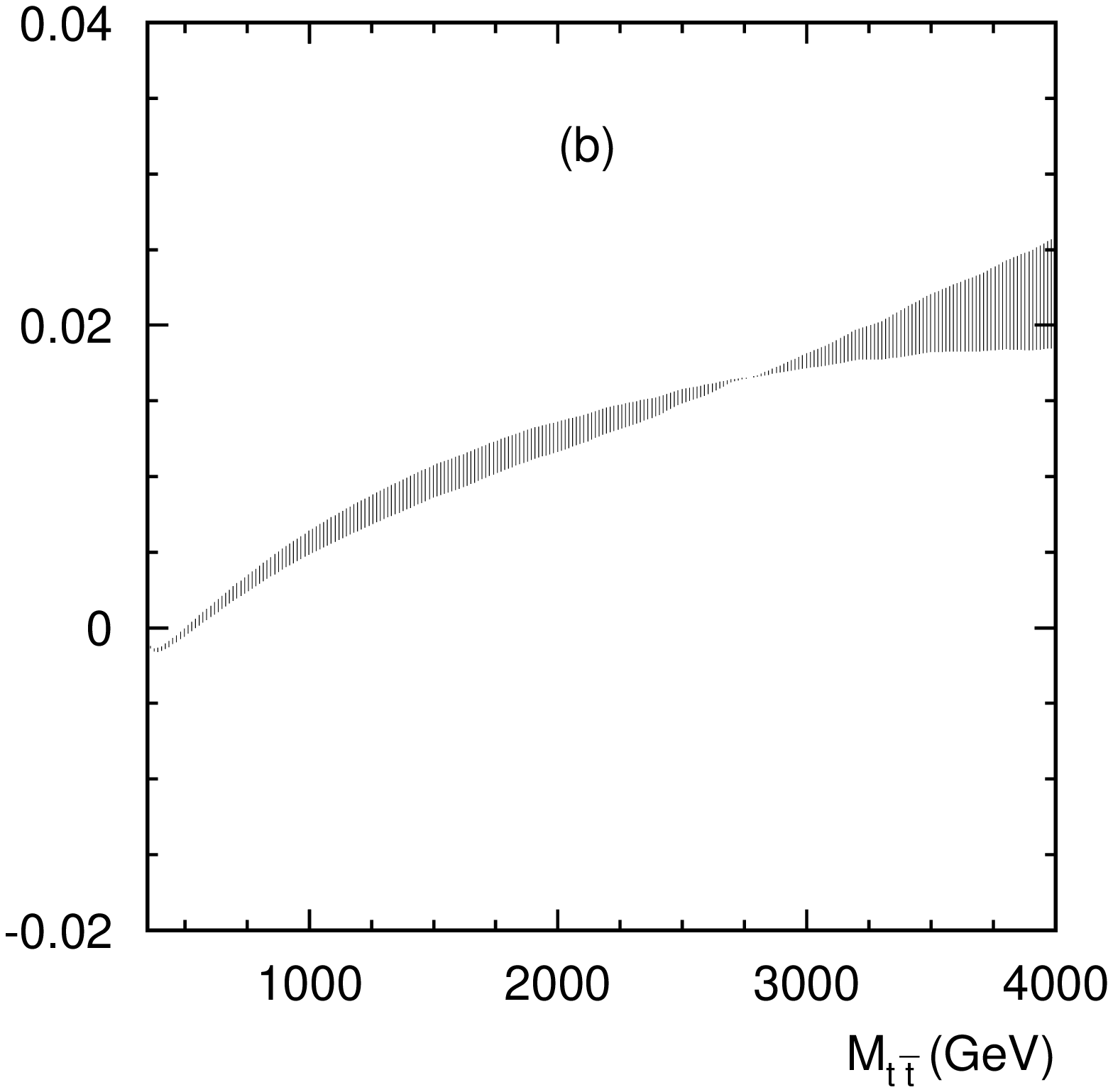}
\end{center}
\caption{
a) Contributions of the various partonic subprocesses to the helicity
asymmetry $\Delta_{hel}$:  initial states
$q \bar q$ $(q\neq b)$ and $gg$ (thin line), $qg$ and ${\bar q}g$
$(q=u,...,b)$ (vertically hatched area), and $b \bar b$ (cross hatched
 area).  b) Sum of the three
contributions shown in a) as computed in \cite{Bernreuther:2008md}.
}\label{fig:dheli}
 \end{figure}
Fig.~\ref{fig:dheli}a displays  the weak-interaction 
induced contributions  i), ii) and iii) 
to $\Delta_{hel}$. (As the SM Yukawa coupling is parity-conserving,
iii) does not depend on $m_H$.) Each correction i) and ii) 
 shows a considerable
scale dependence which, however, cancels to a large extent in the sum
of the two contributions -- c.f. Fig.~\ref{fig:dheli}b.
The corrections i) and ii) reduce the contribution iii) to
$\Delta_{hel}$ by about $50\%$. Thus we find that 
in the SM, $\Delta_{hel} \leq 2\%$  for $\mtt \leq$ 4 TeV.
Selecting events (\ref{ellreact})
 at the LHC with a $t \bar t$ invariant mass larger than some minimum value, we 
 obtain the following SM prediction for the parity-violating asymmetry
 $A_{PV}$:
\begin{equation}
A_{PV}(\mtt>0.5 {\rm TeV}) = 0.0004 , \;  A_{PV}(\mtt>1{\rm TeV}) =
0.0021 , \;
 A_{PV}(\mtt>1.5{\rm TeV}) = 0.005 \, .
\label{apv-resm}
\end{equation}
Such a small effect will hardly be measurable at the
LHC. Nevertheless, 
 the fact that the SM value of $A_{PV}$ is so small, makes
 this observable an
ideal experimental 
sensor for tracing possible new parity-violating interactions
in $t \bar t$ production.
 Thus   $A_{PV}$ should be known  as
precisely as possible within the SM.
On the experimental side, it remains to be investigated 
with which precision  can actually be
measured by an LHC experiment. Given the results from simulations 
\cite{Hubaut:2005er,AguilarSaavedra:2007rs} that
asymmetries related to $W$-boson helicity fractions from top-quark
decay should be measurable at the LHC with an overall uncertainty
 of $\sim 1\%$  ,
 it seems not unrealistic to expect that  $A_{PV}$, which is of a
 similar type as these asymmetries, should eventually be
 measurable with a precision of $\sim 1\%$.
One may also consider PV double spin-asymmetries. However these are
 equivalent, if CP invariance holds,  to the corresponding single-spin
 asymmetries such as $Z_{hel}$, which are experimentally much more
 powerful. These and  other spin 
 issues are discussed in \cite{Bernreuther:2006vg}.

The Standard Model predictions of
Fig.~\ref{fig:dheli}
and of (\ref{apv-resm}) may be used
as reference values in searches for parity-violating effects in
hadronic $t \bar t$ production and decay at the LHC. Which new
 physics effects could possibly lead to  an asymmetry
 $A_{PV}$ at the level of a few percent? Obvious candidates would
be new heavy s-channel resonances that couple to $t\bar t$ pairs
 strongly and in a parity-violating way. 
In two-Higgs doublet or supersymmetric
models the radiative corrections to the $t \bar t$ production
amplitudes can lead to asymmetries larger than those given
   (\ref{apv-resm})   if the new particles are not too heavy 
\cite{Kao:1999kj}.

Finally, a word on how  the weak corrections change in the presence of cuts.
 If one takes into account  only
$t \bar t$ events with  $p_T \geq p_{Tmin}$, the corrections
i), ii)  will not change significantly, as long as  $p_{Tmin}$
is not too large. Choosing, for
instance, $p_{Tmin}=$ 30 GeV does not lead to a significant change 
of the results shown in Figs.~\ref{fig:dpt}~-~\ref{fig:dheli}.  
Eventually, the weak corrections to the distributions discussed here 
should be evaluated  in conjunction 
with the known NLO QCD
corrections, for which the NLO PDF, in particular a
 NLO $b$-quark PDF,  are to be used. 
 The NLO  $b$-quark
PDF enhances the $b$-quark induced
weak contribution to the $\mtt$ distribution and to $\Delta_{hel}$ at
large $\mtt$.

To summarize: distributions and asymmetries
 are key observables in the detailed exploration of
 the dynamics of top quarks at the LHC, which should eventually be
 possible up to energy scales of a few TeV. 
 Therefore these observables should be predicted within the SM as
 precisely as possible. For this reason,  we have analyzed the
 electroweak corrections  to hadronic 
$t\bar{t}$ pair production, and computed the effect
 of these corrections on the  top-quark transverse momentum
 distribution and on the $t\bar{t}$ invariant mass
distribution.  For the LHC these corrections
 are not negligible with respect to the QCD corrections,
 especially at large $p_T$ and  $M_{t\bar{t}}$, respectively.
 Furthermore, we have computed a parity-violating
 forward-backward asymmetry $A_{PV}$, which is induced by the weak interaction.
  The fact that  the SM value of $A_{PV}$ is very small
 makes this observable an ideal tool to search for new PV interactions
 in $t \bar t$ production.

\subsection*{acknowledgments}
W. B. and Z. G. Si  would like to thank Roberto Tenchini and his colleagues
 for having organized this excellent workshop.
This work was supported
by Deutsche Forschungsgemeinschaft (DFG) SFB/TR9, by
DFG-Graduiertenkolleg RWTH Aachen, by NSFC, by NCET, and by 
Huoyingdong Foundation, China.


\begin{thebibliography}{0}


\bibitem{Nason:1987xz}
P.~Nason, S.~Dawson and R.~K.~Ellis,
Nucl.\ Phys.\ B {\bf 303}, 607  (1988),
Nucl.\ Phys.\ B {\bf 327},  49 (1989)
[Erratum-ibid.\ B {\bf 335}, 260 (1990)];
W.~Beenakker, H.~Kuijf, W.~L.~van Neerven and J.~Smith,
Phys.\ Rev.\ D {\bf 40}, 54 (1989);
W.~Beenakker, W.~L.~van Neerven, R.~Meng, G.~A.~Schuler and J.~Smith,
Nucl.\ Phys.\ B {\bf 351},  507 (1991);
  M.~L.~Mangano, P.~Nason and G.~Ridolfi,
  Nucl.\ Phys.\ B {\bf 373}, 295 (1992);
  S.~Frixione, M.~L.~Mangano, P.~Nason and G.~Ridolfi,
  Phys.\ Lett.\ B {\bf 351}, 555 (1995).
%
\bibitem{Bonciani:1998vc}
R.~Bonciani, S.~Catani, M.~L.~Mangano and P.~Nason,
Nucl.\ Phys.\ B {\bf 529},  424 (1998);
N.~Kidonakis, E.~Laenen, S.~Moch and R.~Vogt,
Phys.\ Rev.\ D {\bf 64},  114001 (2001);
  M.~Cacciari, S.~Frixione, M.~L.~Mangano, P.~Nason and G.~Ridolfi,
  JHEP {\bf 0404}, 068 (2004);
  A.~Banfi and E.~Laenen,
  Phys.\ Rev.\ D {\bf 71}, 034003 (2005);
  S.~Moch and P.~Uwer,
  arXiv:0804.1476 [hep-ph].


%
\bibitem{Bernreuther:2000yn}
W.~Bernreuther, A.~Brandenburg and Z.~G.~Si,
Phys.\ Lett.\ B {\bf 483},  99 (2000)
[arXiv:hep-ph/0004184]; 
W.~Bernreuther, A.~Brandenburg, Z.~G.~Si and P.~Uwer,
Phys.\ Lett.\ B {\bf 509},  53 (2001)
[arXiv:hep-ph/0104096].



\bibitem{Bernreuther:2004jv}
  W.~Bernreuther, A.~Brandenburg, Z.~G.~Si and P.~Uwer,
  Nucl.\ Phys.\ B {\bf 690},  81 (2004)
  [arXiv:hep-ph/0403035];
%
W.~Bernreuther, A.~Brandenburg, Z.~G.~Si and P.~Uwer,
Phys.\ Rev.\ Lett.\  {\bf 87}, 242002 (2001) 
[arXiv:hep-ph/0107086].




\bibitem{Melles:2001ye}
  M.~Melles,
  Phys.\ Rept.\  {\bf 375},  219 (2003)
  [arXiv:hep-ph/0104232];
  A.~Denner and S.~Pozzorini,
  Eur.\ Phys.\ J.\ C {\bf 18}, 461 (2001)
  [arXiv:hep-ph/0010201];
 Eur.\ Phys.\ J.\ C {\bf 21}, 63 (2001)
  [arXiv:hep-ph/0104127];
  A.~Denner, B.~Jantzen and S.~Pozzorini,
  arXiv:hep-ph/0608326.



\bibitem{Beenakker:1993yr}
  W.~Beenakker et al.,
  Nucl.\ Phys.\ B {\bf 411},  343 (1994).

\bibitem{Kao:1997bs}
  C.~Kao, G.~A.~Ladinsky and C.~P.~Yuan,
  Int.\ J.\ Mod.\ Phys.\ A {\bf 12},  1341 (1997).


%
\bibitem{Bernreuther:2005is}
  W.~Bernreuther, M.~F\"ucker and Z.~G.~Si,
  Phys.\ Lett.\ B {\bf 633},  54 (2006)
  [arXiv:hep-ph/0508091].     

\bibitem{Kuhn:2005it}
  J.~H.~K\"uhn, A.~Scharf and P.~Uwer,
  Eur.\ Phys.\ J.\ C {\bf 45},  139 (2006)
  [arXiv:hep-ph/0508092].



\bibitem{Bernreuther:2006vg}
  W.~Bernreuther, M.~Fuecker and Z.~G.~Si,
  Phys.\ Rev.\  D {\bf 74}, 113005 (2006);

\bibitem{Bernreuther:2008md}
  W.~Bernreuther, M.~Fuecker and Z.~G.~Si,
  arXiv:0804.1237 [hep-ph], to be published in Phys. Rev. D.


\bibitem{Kuhn:2006vh}
  J.~H.~K\"uhn, A.~Scharf and P.~Uwer,
  Eur.\ Phys.\ J.\  C {\bf 51}, 37 (2007)
  [arXiv:hep-ph/0610335].


\bibitem{Moretti:2006nf}
  S.~Moretti, M.~R.~Nolten and D.~A.~Ross,
  Phys.\ Lett.\  B {\bf 639}, 513(2006) 
  [Erratum-ibid.\  B {\bf 660}, 607 (2008)]
  [arXiv:hep-ph/0603083].



\bibitem{Hollik:2007sw}
  W.~Hollik and M.~Kollar,
  Phys.\ Rev.\  D {\bf 77}, 014008 (2008)
  [arXiv:0708.1697 [hep-ph]].



\bibitem{Kao:1994rn}
  C.~Kao,
  Phys.\ Lett.\ B {\bf 348},  155 (1995)
  [arXiv:hep-ph/9411337].

\bibitem{Kao:1999kj}
  C.~Kao and D.~Wackeroth,
  Phys.\ Rev.\ D {\bf 61}, 055009 (2000)
  [arXiv:hep-ph/9902202].

\bibitem{Beccaria:2005jd}
  M.~Beccaria, F.~M.~Renard and C.~Verzegnassi,
  Phys.\ Rev.\ D {\bf 72}, 093001 (2005) 
  [arXiv:hep-ph/0506230].

\bibitem{Li:1997gh}
  C.~S.~Li, C.~P.~Yuan and H.~Y.~Zhou,
  Phys.\ Lett.\ B {\bf 424},  76 (1998)
  [arXiv:hep-ph/9709275];
  W.~Hollik, W.~M.~M\"osle and D.~Wackeroth,
  Nucl.\ Phys.\ B {\bf 516},  29 (1998)
  [arXiv:hep-ph/9706218];
  M.~Beccaria et al.,
  Phys.\ Rev.\ D {\bf 71},  073003 (2005)
  [arXiv:hep-ph/0412249];
  S.~Berge, W.~Hollik, W.~M.~M\"osle and D.~Wackeroth,
  Phys.\ Rev.\  D {\bf 76}, 034016 (2007)
  [arXiv:hep-ph/0703016];
  D.~A.~Ross and M.~Wiebusch,
  JHEP {\bf 0711}, 041 (2007)
  [arXiv:0707.4402 [hep-ph]].


\bibitem{Aivazis:1993pi}
  M.~A.~G.~Aivazis, J.~C.~Collins, F.~I.~Olness and W.~K.~Tung,
  Phys.\ Rev.\  D {\bf 50}, 3102 (1994)
  [arXiv:hep-ph/9312319].


\bibitem{Pumplin:2002vw}
  J.~Pumplin et al.,
  JHEP {\bf 0207},  012 (2002)
  [arXiv:hep-ph/0201195].


\bibitem{Martin:2004dh}
  A.~D.~Martin, R.~G.~Roberts, W.~J.~Stirling and R.~S.~Thorne,
  Eur.\ Phys.\ J.\  C {\bf 39}, 155 (2005)
  [arXiv:hep-ph/0411040].


\bibitem{Brandenburg:2002xr}
A.~Brandenburg, Z.~G.~Si and P.~Uwer,
Phys.\ Lett.\ B {\bf 539},  235 (2002)
[arXiv:hep-ph/0205023].


%
\bibitem{Hubaut:2005er}
  F.~Hubaut et al.,
  Eur.\ Phys.\ J.\  C {\bf 44S2}, 13 (2005)
  [arXiv:hep-ex/0508061].


\bibitem{AguilarSaavedra:2007rs}
  J.~A.~Aguilar-Saavedra et al.,
  Eur.\ Phys.\ J.\  C {\bf 53}, 689 (2008)
  [arXiv:0705.3041 [hep-ph]].







\end{thebibliography}
\end{document}